%% file: ShapeFaults.tex
\documentclass[conference]{IEEEtran}
\IEEEoverridecommandlockouts
\pdfoutput=1
\usepackage{cite}
\usepackage{amsmath,amssymb,amsfonts}
\usepackage{algorithmic}
\usepackage{graphicx}
\usepackage{textcomp}
\def\BibTeX{{\rm B\kern-.05em{\sc i\kern-.025em b}\kern-.08em
    T\kern-.1667em\lower.7ex\hbox{E}\kern-.125emX}}

\usepackage{listings}
\usepackage{color}
\usepackage{xspace}
\usepackage{xcolor}
\usepackage{float}
\usepackage{booktabs}
\usepackage{stfloats}
\usepackage{url} 
\usepackage{framed} 
\usepackage[T1]{fontenc} 
\usepackage{multirow}
\usepackage[ruled,linesnumbered,noend]{algorithm2e}
\usepackage{threeparttable}
\usepackage{subfigure}
\usepackage{tikz}
\usepackage{mathrsfs}
\usepackage{inconsolata}
\usepackage{pifont}
\usepackage{balance}
\usepackage{verbatim}
\usepackage{color, colortbl}

\usepackage{tikz}

\newcommand{\override}[1]{\textcolor{gray}{#1}}
\newtheorem{definition}{Definition}

\definecolor{dkgreen}{rgb}{0,0.6,0}
\definecolor{gray}{rgb}{0.5,0.5,0.5}
\definecolor{mauve}{rgb}{0.58,0,0.82}

\makeatletter

\newcommand{\Rmnum}[1]{\expandafter\@slowromancap\romannumeral #1@}
\makeatother

\newcommand{\etal}{\hbox{et al.}\xspace}
\newcommand{\eg}{\hbox{e.g.}\xspace}

\newcommand{\vs}{\hbox{vs.}\xspace}

\newcommand\SFData{{SFData}\xspace}

\begin{document}

\title{An Empirical Study on Tensor Shape Faults in Deep Learning Systems}

\author{\IEEEauthorblockN{Dangwei Wu, Beijun Shen, Yuting Chen}
\IEEEauthorblockA{\textit{School of Electronic Information and Electrical Engineering} \\
\textit{Shanghai Jiao Tong University}\\
Shanghai, China \\
\{wudangwei, bjshen, chenyt\}@sjtu.edu.cn}
}

\maketitle

\pagestyle{plain}

\input{0_abstract}

\section{Introduction}
\input{1_introduction}

\section{Background}
\input{2_background}

\section{A Motivating Example}
\input{3_a_motivating_example}

\section{Dataset}
\input{4_dataset}

\section{Analysis and Results}

\input{5_fault_analysis}

\section{Conclusion}
\input{6_conclusion}

\balance
\bibliographystyle{IEEEtran}
\bibliography{ShapeFaults.bib}

\end{document}

%% file: 0_abstract.tex
\begin{abstract}
Software developers frequently adopt deep learning (DL) libraries to incorporate learning solutions into software systems. However, misuses of these libraries can cause various DL faults. Among them, \emph{tensor shape faults} are most prevalent.
Tensor shape faults occur when restriction conditions of operations are not met, leading to many system crashes.
To support efficient detection and fixing of these faults, we conduct an empirical study to obtain a deep insight. We construct \SFData, a set of 146 buggy programs with crashing tensor shape faults (i.e., those causing programs to crash). By analyzing the faults in \SFData, we categorize them into four types and get some valuable observations. 
\end{abstract}

\begin{IEEEkeywords}
Tensor shape fault, crash, deep learning systems
\end{IEEEkeywords}

%% file: 1_introduction.tex
Deep learning (DL) provides efficient solutions for a number of problems that were difficult to solve with traditional
computing techniques; \eg, automatic driving, speech recognition and machine translation.
They are widely supported by DL libraries, such as TensorFlow~\cite{TensorFlow}, Keras~\cite{Keras}, Torch~\cite{torch}, Cafe~\cite{Cafe} and Theano~\cite{Theano}; these libraries facilitate programmers to develop, train, and run models efficiently.
While DL libraries make it easy to incorporate DL into software systems, engineers still fail in using them correctly because of their unique semantic and data requirements~\cite{wanmachine}.

Tensor shape faults are the main type of DL faults; they can occur in all stages of a DL pipeline~\cite{Islam2019ACS}.
DL programs build models on the basis of computational graphs: each node in a computational graph represents an operation (OP), and a tensor (i.e., an input or output of an operation) corresponds to an edge of the graph. Each tensor has its shape properties, such as the number of dimensions and the size of each dimension. When tensors flow and change in computational graphs, their shapes also flow and change.
An OP manipulates tensors and usually restricts their shapes.
When a restriction condition of OPs is not met, a \emph{tensor shape fault} is produced, which usually leads to a program crash.
For example, a misuse of a DL API in a two-dimensional tensor multiplication may cause a tensor shape fault, because the multiplication requires the size of the second dimension of the first matrix to be equal to the size of the first dimension of the second matrix. Fig. \ref{example} shows a more complex example (\S 3).

A recent study reveals that tensor shape faults are frequent, covering 45\% of failures in TensorFlow programs~\cite{Verma2020ShapeFlowDS}. The high prevalence of tensor shape faults and their serious consequences
inspires research on shape faults.
Previous efforts focus on detecting  tensor shape faults.
One solution to this is \emph{static shape fault detection}, which analyzes the source code of DL programs against detection rules.
Ariadne~\cite{Dolby2018AriadneAF} tracks and analyzes tensor usages, checking whether each tensor is provided with the desired shape.
Pythia~\cite{Lagouvardos2020StaticAO} extends Ariadne, which translates Python programs into intermediate representations and designs rules to detect shape faults w.r.t. TensorFlow's OPs.
All the static techniques require detection rules to be well prepared.
Another mainstream is \emph{dynamic shape fault detection}, which analyzes the programs' execution information and detects tensor shape faults.
The only previous work is ShapeFlow~\cite{Verma2020ShapeFlowDS}, an abstract interpreter for TensorFlow programs.
It rewrites TensorFlow's APIs and runs programs, extracting the tensor shapes and then constructing shape computational graphs for tracking shape changes.
Up to now, there exist few researches on automatically repairing these faults.

To detect and repair tensor shape faults efficiently,
a dataset of shape faults should be constructed from real projects, and analyzed to obtain insights into tensor shape faults.
Zhang \etal~\cite{Zhang2018AnES} build a dataset of buggy TensorFlow programs in their empirical study, while the dataset contains only 14 tensor shape faults collected from StackOverflow and 9 ones from Github. Islam \etal construct a dataset from StackOverflow~\cite{Islam2019ACS}, containing 32 and 37 tensor shape faults in TensorFlow and Keras programs, respectively. Meanwhile, Islam's dataset does not contain buggy programs and their patches, since it is not designed for program analysis and bug repair.

In this paper, we empirically study crashing tensor shape faults in StackOverflow. We build \SFData, a dataset of crashing tensor shape faults that contains 146 buggy programs and their crash messages, patches and test data.
By analyzing these data, we reveal four types of faults, helping researchers understand these faults and as well engineers detect and repair them.
We have released our whole data suite of \SFData and detailed study results on Github~\footnote{https://github.com/tensfa/tensfa}.

The rest of the paper is organized as follows. Section II provides
a background and related work of tensor shape faults.
Section III gives a motivating example.
Section IV describes the construction process of \SFData.
Section V presents the results of data analysis on \SFData,
and Section VI concludes.

%% file: 2_background.tex
\subsection{Preliminaries}
\begin{figure}[t]
\centering
\includegraphics[width=0.4\textwidth]{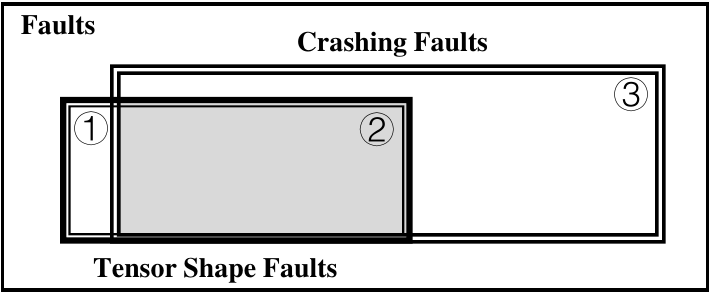}
\caption{Relationship between tensor shape faults and crashing faults. Here \ding {172},  \ding {173},  and \ding {174} represent non-crashing tensor shape faults,  crashing tensor shape faults, and the other crashing faults, respectively. In our study, we mainly focus on crashing tensor shape faults.}
\label{faults}
\end{figure}

First defines the notions used in this paper. 

\begin{definition}
\textit{(Operation)}. An operation $OP$ is a function that accepts a set of parameters, each of which can either be an input tensor or a variable value, and produces a set of output tensors and may as well updates the parameters' values:
$$
(ot_{0}, ot_{1}, \dots, ot_{m}~|~ov_{0}, ov_{1}, \dots, ov_{n}) $$
$$= OP (it_{0}, it_{1}, \dots, it_{p}~|~iv_{0}, iv_{1}, \dots, iv_{q}),$$
where $ot_{0 \leq i  \leq  m}$ and $it_{0 \leq i  \leq  n}$ are an output tensor and an input one, respectively; $ov_{0 \leq i  \leq  p}$ and $iv_{0 \leq i  \leq  q}$ is a left value and a right value of the operation.
\end{definition}

\smallskip

\begin{definition}
\textit{(Shape Restriction)}. Let $t$ be a tensor and $shape(t)$ be the dimensionality of $t$. A shape restriction is a set of conditions a tensor's shape needs to meet:
$$
restrict(t)=\{cond_1,cond_2, \dots,cond_n\};$$
$OP$'s shape restriction is a set of the restrictions on all of its tensors, i.e.,
$$restrict(OP)=(\cup_{0\leq k \leq m} restrict(ot_k)) $$
$$\ \ \ \ \ \ \ \ \ \ \ \ \ \ \ \
\cup~(\cup_{0\leq k \leq p} restrict(it_k)).$$
\end{definition}


\smallskip

\begin{definition}
\textit{(Tensor Shape Fault)}. 
A tensor shape fault in a DL program is a violation of shape restrictions, i.e.,
$$tsf(OP) \implies (\exists t \in \{it_0, \dots, it_p, ot_0, \dots, ot_m\}, $$
$$\exists cond \in restrict(t)|\bullet ~cond(shape(t))= false).$$
\end{definition}

\medskip
As Fig. \ref{faults} shows, a tensor shape fault may or may not cause a crash. On the other hand, a crash may or may not be raised by a tensor shape fault. Correspondingly, we define \textit{crashing tensor shape fault} as follows.

\smallskip
\begin{definition}
\textit{(Crashing Tensor Shape Fault)}. A \emph{crashing fault} is a fault that can cause a program to crash. A \emph{crashing tensor shape fault} is a tensor shape fault that can cause a DL program to crash.
\end{definition}

\subsection{Related Work}
There has been a number of researches~\cite{Zhang2018AnES, Islam2019ACS, Islam2020RepairingDN, HumbatovaJBR0T20, jia2020empirical}  on studying DL faults from open source repositories.
Humbatova \etal~\cite{HumbatovaJBR0T20} introduce a taxonomy of faults in DL systems.
Li \etal~\cite{jia2020empirical} find that many TensorFlow bugs are in its interfaces.
Zhang \etal~\cite{Zhang2018AnES} collect data from GitHub and StackOverflow to study faults in TensorFlow programs, examining root causes and symptoms of the faults as well as methods used for detecting and locating faults. Their dataset contains 23 TensorFlow programs with shape faults.
Islam \etal~\cite{Islam2019ACS} examine programs related to five popular deep learning libraries 
to understand types of errors, root causes, impacts, and anti-patterns. They further study fault repairs on GitHub and StackOverflow and summarize the repair patterns and challenges~\cite{Islam2020RepairingDN}. Islam's dataset contains 69 TensorFlow/Keras programs with shape faults.

Compared with the above studies, our empirical study and dataset focus on crashing tensor shape faults, while previous approaches are targeting general DL faults.
Furthermore, \SFData contains the shape faults and their buggy programs, patches and tests, which are critical to reproduce and repair faults. Besides, \SFData is  larger than existing datasets.

%% file: 3_a_motivating_example.tex
\newbox\ExampleBox
\begin{lrbox}{\ExampleBox}
\begin{lstlisting}
import tensorflow as tf
import numpy as np
train_n = 10
X_training = np.array(np.random.random(|{\quad\quad}| (train_n, 100, 100, 3)), dtype=np.float32) #
Y_training = np.eye(2)[np.random.randint(0, 2, train_n, dtype=np.int32)] #
input = 100*100*3
X = |\override{tf.placeholder}|(tf.float32, [1, 100, 100, 3]) #
W = |\override{tf.Variable}|(|\override{tf.zeros}|([input, 2])) #
b = |\override{tf.Variable}|(|\override{tf.zeros}|([2])) #
init = |\override{tf.global\_variables\_initializer}|()
Y = |\override{tf.nn.softmax}|(|\override{tf.matmul}|(|\override{tf.reshape}|(X, [-1, input]), W) |\override{+}| b) #
Y_ = |\override{tf.placeholder}|(tf.float32, [None, 2]) #
cross_entropy = |\override{-tf.reduce\_sum}|(Y_ |\override{*}| |\override{tf.log}|(Y)) #
is_correct = |\override{tf.equal}|(|\override{tf.argmax}|(Y,1), |\override{tf.argmax}|(Y_,1)) #
accuracy = |\override{tf.reduce\_mean}|(|\override{tf.cast}|(|{\quad\quad}| is_correct, tf.float32)) #
optimizer = |\override{tf.train.GradientDescentOptimizer}|(0.003)
train_step = |\override{optimizer.minimize}|(cross_entropy)
sess = |\override{tf.Session}|()
|\override{sess.run}|(init)
for i in range(len(X_training)):
   batch_X, batch_Y = X_training[i], Y_training[i] #
|\textcolor[RGB]{0,100,0}{+\quad \# Patch1}|
|\textcolor[RGB]{0,100,0}{+\quad batch\_x = np.expand\_dims(batch\_X, 0) \# fix the code using an FDS pattern}|
|\textcolor[RGB]{0,100,0}{+\quad batch\_Y = np.expand\_dims(batch\_Y, 0) \# fix the code using an LDS pattern}|
   train_data={X: batch_X, Y_: batch_Y}
   |\override{sess.run}||\textcolor{red}{(train\_step,feed\_dict=train\_data)}| #
   a,c = |\override{sess.run}|([accuracy, cross_entropy], feed_dict=train_data) #
\end{lstlisting}
\end{lrbox}

\begin{figure}[t]
\centering
\subfigure[A buggy program \texttt{BiClassifier} and a patch to the program. Here we call the patched program \texttt{BiClassifier'}.]
{
\begin{minipage}[b]{0.5\textwidth}
\usebox\ExampleBox
\vspace{0.1cm}
\end{minipage}
}

\subfigure[Computational graph of the \texttt{BiClassifier} program.]{
\begin{minipage}[b]{0.5\textwidth}
\includegraphics[width=0.98\textwidth]{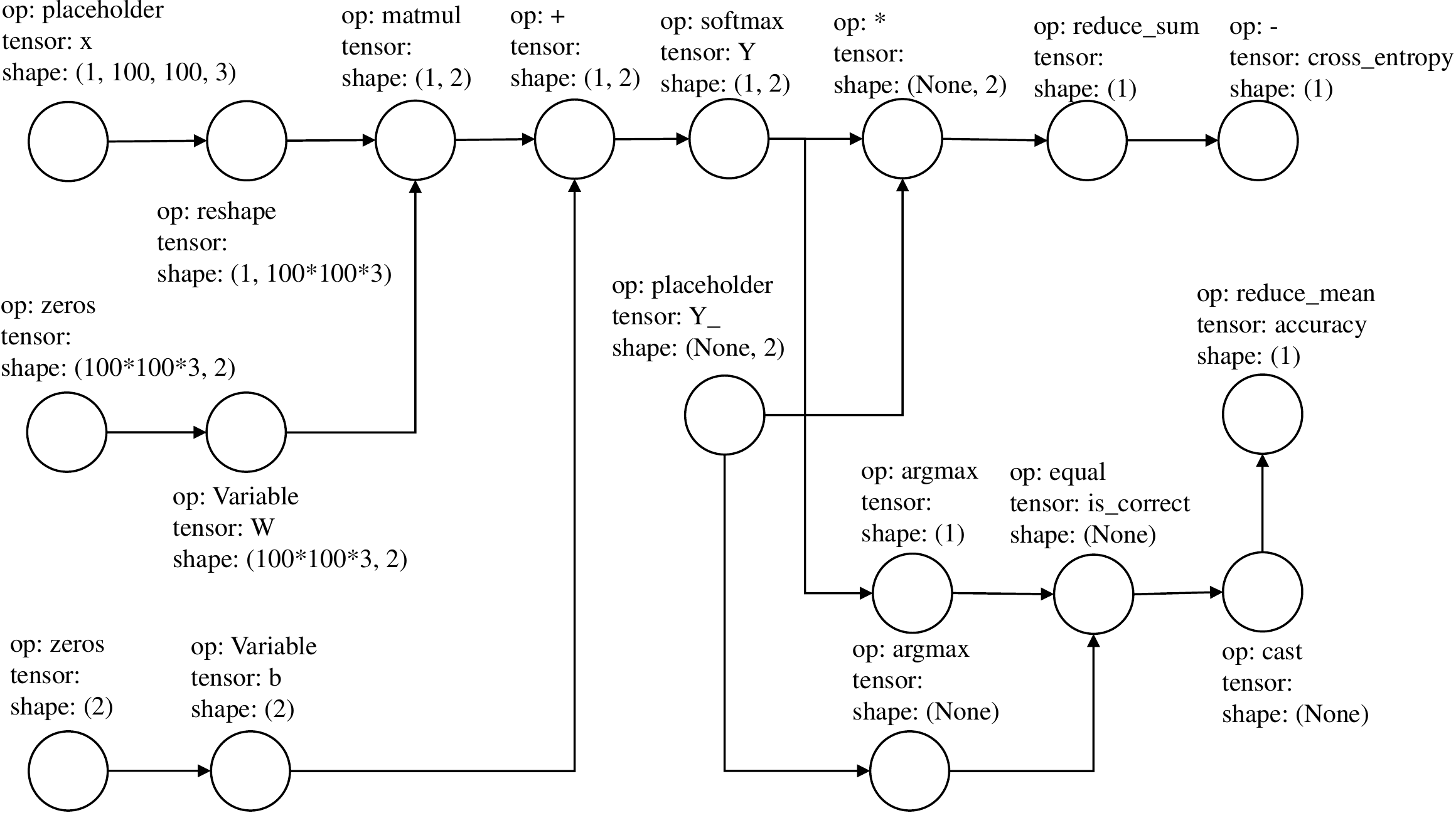}
\end{minipage}
}
\caption{An example in \SFData. This example is collected from StackOverflow \#55237206.}
\label{example}
\end{figure}

We next illustrate tensor shape faults in DL programs using an example.
Fig. \ref{example}(a) shows a buggy program collected from StackOverflow \#55237206. The \texttt{BiClassifier} program implements a linear binary classifier. In its code, the operator at line 26 triggers a crash, and the lines in green (lines 22$\sim$24) stand for a repair patch.

The program's computational graph is shown in Fig. \ref{example}(b).  During its construction stage, the program runs normally. However, during its execution stage, the fault occurs, resulting in a program crash. When the data is loading into the graph and the forward computation is performed, the program triggers a crash ``\texttt{ValueError: Cannot feed value of shape (100, 100, 3) for Tensor Placeholder:0, which has shape (1, 100, 100, 3)}'' at line 26.
The message indicates that there exists an incompatibility between the shapes \texttt{(100, 100, 3)} and \texttt{(1, 100, 100, 3)}. The root cause is that the feature data and the label data lack their batch size dimensions.

%% file: 4_dataset.tex
\begin{table}[t]
\centering
\caption{Statistics of IslamData and \SFData.}
\setlength\tabcolsep{2.4pt}
\label{table2}
\begin{tabular}{ccccc}
\hline
\multicolumn{1}{c|}{}                                      & \multicolumn{2}{c|}{\textbf{IslamData}}                                  & \multicolumn{2}{c}{\textbf{SFData}}                       \\
\multicolumn{1}{c|}{}                                      & \multicolumn{2}{c|}{\textbf{\# of Programs (\# for repair)}}               & \multicolumn{2}{c}{\textbf{\# of Programs (\# for repair)}} \\ \cline{2-5} 
\multicolumn{1}{c|}{\multirow{-3}{*}{\textbf{Fault Type}}} & \textbf{TensorFlow} & \multicolumn{1}{c|}{\textbf{Keras}}                & \textbf{TensoFlow}             & \textbf{Keras}            \\ \hline
\rowcolor[HTML]{EFEFEF} 
\multicolumn{5}{c}{\cellcolor[HTML]{EFEFEF}Crashing Tensor Shape Faults (positive samples)}                                                                                                                 \\ \hline
\multicolumn{1}{c|}{PRI}                                   & 19(0)               & \multicolumn{1}{c|}{3(0)}                          & 26(26)                         & 6(6)                      \\
\multicolumn{1}{c|}{ALI}                                   & 4(0)                & \multicolumn{1}{c|}{5(0)}                          & 6(6)                           & 16(16)                    \\
\multicolumn{1}{c|}{FII}                                   & 4(0)                & \multicolumn{1}{c|}{24(0)}                         & 15(15)                         & 42(42)                    \\
\multicolumn{1}{c|}{LOI}                                   & 5(0)                & \multicolumn{1}{c|}{5(0)}                          & 16(16)                         & 31(31)                    \\
\multicolumn{1}{c|}{Subtotal}                              & 32(0)               & \multicolumn{1}{c|}{37(0)}                         & 63(63)                         & 95(95)                    \\ \hline
\rowcolor[HTML]{EFEFEF} 
\multicolumn{5}{c}{\cellcolor[HTML]{EFEFEF}Other Crashing Faults (negative samples)}                                                                                                                  \\ \hline
\multicolumn{1}{c|}{Other}                                 & 63(0)               & \multicolumn{1}{c|}{43(0)}                         & 86(0)                          & 98(0)                     \\ \hline
\rowcolor[HTML]{EFEFEF} 
\multicolumn{1}{c|}{\cellcolor[HTML]{EFEFEF}Total}         & 95(0)               & \multicolumn{1}{c|}{\cellcolor[HTML]{EFEFEF}80(0)} & 149(63)                        & 193(95)                   \\ \hline
\end{tabular}
\end{table}

We construct a set of buggy programs with crashing tensor shape faults, \SFData, 
by taking the following steps: collecting shape-fault-related posts in StackOverflow, analyzing these faults and further associating them with other supplementary information (e.g., crash messages, source code, patches and test data).

\subsection{Data collection}
On StackOverflow, we search for posts using keywords ``\texttt{[tensorflow] or [keras] ValueError shape answers:1..}'', where 
\begin{itemize}
\item \texttt{TensorFlow} and \texttt{Keras} programs are chosen because they are two popular, and as well closely-related deep learning libraries. They are open-sourced at GitHub, where TensorFlow is of 154K stars and 84.3K forks, and Keras is of 50.9K stars and 18.7K forks, as of Apr. 2021;

\item the keyword \texttt{ValueError} indicates that posts related with ValueError are acquired, since an empirical study reveals that 71.43\% of tensor shape faults in TensorFlow and Keras programs are reported as value errors \cite{Islam2019ACS};

\item and the keyword \texttt{answers:1..} indicates that we search for posts with at least one answer, aiming at improving the quality of the obtained posts. 
\end{itemize}

\subsection{Data processing}
We manually check the retrieved posts, comments and answers to determine whether the corresponding programs contain shape faults and eliminate the duplicate ones. Four  strategies are then taken 
for processing data, allowing a high-quality dataset to be constructed.

\textit{S1: Extracting information from posts.} For each post, we extract the buggy program, the crash message, the patch and the tests if they are available; we also extract the post descriptions, comments, and answers.

\textit{S2: Reusing code in Github.} We search for open source projects on Github to retrieve the code, bug reports, patches, and/or test data if a fault is incompletely described (e.g., the patch is missing).

\textit{S3: Generating random tests.} For the  programs without tests, we generate random tests as their inputs; the random strategy  is frequently used in answering issues on StackOverflow, such as the post \#58277932.

\textit{S4: Producing crash messages.} For each buggy program, we run it and its patch against the tests, obtaining the crash message and validating the patch. 

All of the posts, buggy programs, faults, and patches are carefully checked. In this study, we check the data manually, but discuss and confirm the results by taking a rigorous code/fault inspection process.

\subsection{A summary of the \SFData dataset}
As Table \ref{table2} shows,
\SFData consists of 146 buggy programs with crashing tensor shape faults. Among them, 59 are TensorFlow programs and 87 are Keras ones. Each record has the attributes: \texttt{[ProgramID, crash message, \{fault type\}, postID in StackOverflow, original code, patch, \{test data\}]}; for each  tensor shape fault, its program, tests and patch are provided such that the fault can be reproduced.

Note that the average number of faults in each program is 1.08 (158/146), because a buggy program can contain one or more shape faults.
Besides, we provide other crashing faults as negative samples; they are collected from StackOverflow and Zhang's empirical study~\cite{Zhang2018AnES}.

\medskip\noindent\textbf{\SFData \vs IslamData.} IslamData is a dataset whose buggy programs with tensor shape faults are extracted from Islam \etal's empirical study~\cite{Islam2019ACS}. It can be used to measure the fault detection techniques. Comparatively, \SFData contains the source code of buggy programs, crash messages, (expected) patches and tests; it is also designed for measuring fault repair techniques, letting the patches be the ground truths of the repair techniques. 

%% file: 5_fault_analysis.tex
\begin{table}[t]
\centering
\begin{threeparttable}[b]
\caption{Fault types of crashing tensor shape faults in DL programs.}
\setlength\tabcolsep{4pt}
\label{table1}
\begin{tabular}{c|ll}
\hline
\textbf{Stage*}               & \multicolumn{1}{c}{\textbf{Type}}                                               & \multicolumn{1}{c}{\textbf{Description}}                                                                     \\ \hline
\multirow{2}{*}{Construction} & \begin{tabular}[c]{@{}l@{}}PRI\end{tabular} & \begin{tabular}[c]{@{}l@{}}Parameter Restriction Incompatible: OP parameter\\ shapes and OP restrictions are not compatible.\end{tabular}        \\
                              & \begin{tabular}[c]{@{}l@{}}ALI\end{tabular}         & \begin{tabular}[c]{@{}l@{}}Adjacent Layer Incompatible: Shapes of adjacent\\ layers are not compatible.\end{tabular}                      \\ \hline
\multirow{2}{*}{Execution}    & \begin{tabular}[c]{@{}l@{}}FII\end{tabular}          & \begin{tabular}[c]{@{}l@{}}Feature Input Incompatible: Feature data shapes\\ and the model input shapes are not compatible.\end{tabular} \\
                              & \begin{tabular}[c]{@{}l@{}}LOI\end{tabular}           & \begin{tabular}[c]{@{}l@{}}Label Output Incompatible: Label data shapes\\ and the model output shapes are not compatible.\end{tabular}  \\ \hline
\end{tabular}

\begin{tablenotes}
 \item[*] ``Stage'' refers to a stage of constructing or executing a computational\\ graph.
\end{tablenotes}
\end{threeparttable}
\end{table}

\subsection{Shape Fault Analysis}
After an analysis of the \SFData dataset, we categorize all of the crashing tensor shape faults into four types, as Table \ref{table1} shows. Let a computational graph be constructed for deep learning. A shape fault may occur during either its construction or its execution stage. 

\begin{itemize}
    \item 
The \emph{construction-stage faults} can be divided into {PRI}  and {ALI}. A PRI fault occurs when the shapes of parameters of a general OP violate the shape  restrictions (\eg,  StackOverflow \#54452974). An ALI fault occurs when the output shape of the previous layer is incompatible with the input shape of the next layer (\eg,  StackOverflow \#47842931).

\item The \emph{execution-stage faults} can be divided into {FII}  and {LOI}. An FII fault refers to an incompatibility between a feature data shape and an input tensor's shape. An LOI fault refers to an incompatibility between a label data shape and an output tensor's shape (\eg,  StackOverflow \#42821125). 

\end{itemize}

Statistically, the execution-stage faults account for 65.82\% of crashing tensor shape faults. The construction-stage faults account for 23.16\% and 50.79\% of crashing tensor shape faults in Keras and TensorFlow programs, respectively. One main reason for this is that Keras's APIs are well encapsulated and easy to use, reducing tensor shape faults during the graphs' construction stage.

\begin{framed}
\noindent\emph{Finding 1}:
Shape faults may occur during the construction and the execution of computational graphs, while the execution-stage faults are more than the construction-stage ones. The better a DL library encapsulates its APIs, the less possible its programs suffer from tensor shape faults during the construction stage. 
\end{framed}

\subsection{Analysis of Fault Detection Capability}
We conduct further analysis on \SFData by employing existing fault detection techniques. 
We observe that Ariadne~\cite{Dolby2018AriadneAF}, Pythia~\cite{Lagouvardos2020StaticAO} and ShapeFlow~\cite{Verma2020ShapeFlowDS}  require a set of well-designed checking rules for detecting tensor shape faults. 
However, there are 3016 OPs in TensorFlow and 535 OPs in Keras, and each OP may have trivial restrictions on parameters. 
Correspondingly, all of these approaches are less effective, as they define only a small number of restriction rules; restrictions may also change over time, as the DL libraries can update frequently. Furthermore, the rule-based approaches are facing the overfitting problem, leading to a low recall in shape fault detection.

Notably, although restrictions are diverse, the crash messages raised by tensor shape faults are similar to each other. 
For example, 91.78\% of crashing tensor shape faults produce shape-related crash messages (i.e., messages containing the word ``shape''). Meanwhile, only 11.80\% of other crashing faults can produce such messages.
It is promising to leverage ML techniques to learn from crash messages for distinguishing tensor shape faults from other crashing faults. 

\begin{framed}
\noindent\emph{Finding 2}:
Rule-based approaches suffer from the overfitting problem,  making their detection capabilities be constrained by the sufficiency of the rules; existing techniques also face challenges in adapting to the evolution of DL libraries. Correspondingly, detection techniques may benefit from machine learning methods in classifying tensor shape faults and exploring fault detecting rules.
\end{framed}

%% file: 6_conclusion.tex
In this paper, we conduct an empirical study on crashing tensor shape faults. It constructs a dataset that contains 146 buggy programs and their faults, patches and test data, and reveals four types of faults and some key observations.

In the future, we plan to extend our \SFData via extracting tensor shape faults from Github and analyzing buggy programs on other DL libraries. We will also explore new detection and repair techniques of tensor shape faults, using the observations and \SFData.